\documentclass{sig-alternate-05-2015}

\usepackage{amsmath}
\usepackage{ctable}
\usepackage{multirow}
\usepackage[caption=false]{subfig}
\usepackage{graphicx}
\usepackage{rotating}
\usepackage{enumerate}
\usepackage{balance}

\newcounter{lizcounter}
\DeclareRobustCommand{\liz}[1]{\textbf{/* #1 (liz) */}\stepcounter{lizcounter}\typeout{LaTeX Warning: liz comment \thelizcounter: #1 (line \the\inputlineno)}}

\begin{document}

\title{The Influence of Frequency, Recency and Semantic Context on the Reuse of Tags in Social Tagging Systems}

\numberofauthors{2}
\author{
\alignauthor
Dominik Kowald\\
       \affaddr{Know-Center}\\
			 \affaddr{Graz University of Technology}\\
       \affaddr{Graz, Austria}\\
       \email{dkowald@know-center.at}		
\alignauthor
Elisabeth Lex\\
       \affaddr{Knowledge Technologies Institute}\\
			 \affaddr{Graz University of Technology}\\
       \affaddr{Graz, Austria}\\
       \email{elisabeth.lex@tugraz.at}
}


\maketitle

\begin{abstract}
In this paper, we study factors that influence tag reuse behavior in social tagging systems. Our work is guided by the activation equation of the cognitive model ACT-R, which states that the usefulness of information in human memory depends on the three factors usage frequency, recency and semantic context. It is our aim to shed light on the influence of these factors on tag reuse. In our experiments, we utilize six datasets from the social tagging systems Flickr, CiteULike, BibSonomy, Delicious, LastFM and MovieLens, covering a range of various tagging settings. Our results confirm that frequency, recency and semantic context positively influence the reuse probability of tags. However, the extent to which each factor individually influences tag reuse strongly depends on the \emph{type of folksonomy} present in a social tagging system. Our work can serve as guideline for researchers and developers of tag-based recommender systems when designing algorithms for social tagging environments.
\end{abstract}



\section{Introduction} \label{sec:intro}
With the advent of the Web 2.0, social tagging has become an essential tool to collaboratively annotate content with freely chosen \emph{tags} (i.e., keywords). The result of a social tagging process is a network connecting users, content resources and tags, which is referred to as \emph{folksonomy} \cite{helic2012navigational}. Since tags can be used to search for, browse and share content, they facilitate discovery and navigation in the Social Web \cite{korner2010stop}. Many social networks such as Medium, Twitter, Instagram and Facebook have adopted tagging in the form of hashtags as well \cite{romero2011differences}. Therefore, previous work has proposed tag recommendation algorithms with the aim to assist users in finding descriptive tags \cite{dellschaft2012measuring} and to control the shared tag vocabulary \cite{wagner2014semantic}. In this respect, recent research has shown that a substantial amount of tag assignments can be explained by analyzing the information access in human memory \cite{fu2008microstructures,seitlinger2014verbatim}, which is mainly influenced by past usage frequency, recency and the current semantic context \cite{anderson1991reflections,anderson2004integrated}.

Even though there is already a large body of research available, which proposes algorithms for recommending tags \cite{jaschke2007tag,marinho2008collaborative,hotho2006folkrank,krestel2009latent,zhang2012integrating,rendle2010pairwise}, none of these approaches incorporate all three aforementioned factors. For example, FolkRank \cite{hotho2006folkrank} solely utilizes tag frequency and the current semantic context, whereas GIRP \cite{zhang2012integrating} solely builds on tag frequency and recency. Typically these methods have been evaluated as integrated models and thus, it remains unclear to what extent the individual factors of the approaches contribute to the final algorithmic performance.

\vspace{2mm} \noindent \textbf{The present work: Factors that influence tag reuse.} In this work, we study factors that potentially influence the tag reuse behavior in social tagging systems. Specifically, we analyze the influence of frequency, recency and semantic context on the reuse of tags. Hence, it is our goal to better understand to what extent these factors can be exploited to predict the reuse of tags given a specific \emph{folksonomy type}. This should lead to a guideline for designing and implementing tag prediction algorithms for given environments.

To that end, we integrate and extend our previous work on tag recommender systems \cite{www_bll,springer_bllac}, where we adapted the activation equation from the cognitive architecture ACT-R \cite{anderson2004integrated} to develop a model termed \emph{BLL$_{AC}$}. This model enables the prediction of future tag assignments for a user $u$ by modeling the usefulness of a piece of information $i$ -- in our case, a tag -- in $u$'s memory based on three factors: (i) how \textit{frequent} $i$ was used by $u$ in the past, (ii) how \textit{recent} (i.e., the time since the last usage) $i$ was used by $u$ in the past, and (iii) how useful $i$ is for $u$ in the current \textit{semantic context}. 

This is achieved by the two components of the activation equation: First, the base-level learning component (\emph{BLL}), which integrates the factors of frequency and recency via a power function for reflecting the time-depended decay of tag reuse \cite{anderson1991reflections} and second, the associated component (\emph{AC}), which models the current semantic context as the similarity of tag $i$ to tags already associated with the currently tagged resource $r$. However, since BLL$_{AC}$ utilizes the three factors, (i) frequency, (ii) recency, and (iii) semantic context, as an integrated model, it is still unclear to what extent these factors individually contribute to the efficacy of the model. Besides, we assume that the influence of these factors on predicting tag reuse depends on the folksonomy type, i.e., \emph{narrow} (e.g., Flickr), \emph{mixed}\footnote{With \emph{mixed} folksonomies, we denote folksonomies that cannot strictly be categorized into the narrow or broad setting.} (e.g., BibSonomy) or \emph{broad} (e.g., MovieLens) \cite{helic2012navigational}, of the given social tagging system. This leads to the following two research questions of our work:
\begin{itemize}
\item \textit{RQ1:} How are the factors of frequency, recency and semantic context influencing a tag's probability of being reused in social tagging systems?
\item \textit{RQ2:} Can the factors of frequency, recency and semantic context be exploited to efficiently predict a user's tag reuse given a specific folksonomy type?
\end{itemize}
\noindent \textbf{Methods and findings.} In order to address \textit{RQ1}, we conducted an empirical study on six social tagging datasets (i.e., Flickr, CiteULike, BibSonomy, Delicious, LastFM and MovieLens), in which we analyzed the influence of the three factors frequency, recency and semantic context on the reuse probability of tags (see Section \ref{sec:emp}). Next, to answer \textit{RQ2}, we carried out a prediction study on the same datasets, in which we not only compared algorithms that reflect these factors individually but also approaches that combine these factors or incorporate social influences, e.g., by suggesting related tags of other users (see Section \ref{sec:pred}). We find that frequency, recency and semantic context positively influence the reuse probability of tags in all systems (\textit{RQ1}) and that the efficacy of these factors for predicting a user's tags depends on the folksonomy type of the given system (\textit{RQ2}).

To the best of our knowledge, this is the first study of its kind, which analyzes the reuse of tags to provide a transparent overview of the factors that influence the prediction of tags and, at the same time, relates these factors to the folksonomy type of the given system. Our findings may serve as a guideline for researchers and developers of tag-based recommender systems with regard to choosing the right prediction and recommendation methods for given environments.

\section{Empirical Study} \label{sec:emp}
In this section, we present the datasets, methodology and results of our empirical study carried out to address \textit{RQ1}. %

\subsection{Datasets} \label{sec:data}
For the sake of this study, we turn to publicly available, real-world datasets gathered from the six social tagging systems Flickr\footnote{\url{https://www.uni-koblenz.de/FB4/Institutes/IFI/AGStaab/Research/DataSets/PINTSExperimentsDataSets}}, CiteULike\footnote{\url{http://www.citeulike.org/faq/data.adp}}, BibSonomy\footnote{\url{http://www.kde.cs.uni-kassel.de/bibsonomy/dumps}}, Delicious\footnote{Same as Flickr.}, LastFM\footnote{\url{http://grouplens.org/datasets/}} and MovieLens\footnote{Same as LastFM.}. To make our results comparable and to ensure reproducibility, we utilize the exact same dataset samples that were used in the study of \cite{recsys_eval}. A major advantage of these dataset samples is that they were created without any $p$-core pruning technique to ensure an unbiased evaluation \cite{doerfel2013analysis}. Additionally, these datasets represent social tagging systems of various domains (i.e., images, Web links, references, music and movies), and differ in size and narrowness degree. The narrowness degree is defined as the average number of posts per resource, which allows to distinguish between narrow (Flickr), mixed (CiteULike, BibSonomy, Delicious) and broad (LastFM, MovieLens) folksonomies \cite{helic2012navigational}. The final statistics of our datasets are shown in Table \ref{tab:datasets}.
 
\subsection{Methodology} \label{sec:emp_meth}
In order to analyze the tag reuse behavior of users, we split our datasets into training and test sets via an evaluation protocol, which persists the chronological order of the data. Thus, for each user $u$, we sorted his/her $n$ posts by time, allocated the $n^{th}$ (i.e., the most recent) post to the test set and the first $n - 1$ posts to the training set \cite{recsys_eval,jaschke2007tag}.

Next, in order to quantify the influence of frequency, recency and semantic context on the reuse of tags, we compared the tag assignments of the first $n - 1$ posts in the training set (i.e., reflecting the past) of user $u$ with the tag assignments of $u$'s $n^{th}$ post in the test set (i.e., reflecting the future). More specifically, for each tag $i$ of $u$, we counted the number of times $i$ was used by $u$ in the training set and determined if $i$ was also reused by $u$ in the test set. Finally, to obtain a statistically reliable value, we pooled together the tags of all users with the same frequency value and calculated the proportion of reused tags to determine the influence of tag frequency on the reuse probability.

We followed a similar procedure to study the influence of recency and semantic context. In the case of tag recency, we calculated the days elapsed since the last use of $i$ by $u$. In the case of the semantic context, we determined the tag co-occurrence value between $i$ and the tags already assigned to the currently tagged resource $r$ (i.e., the second component of the activation equation). Then, as in the case of tag frequency, we pooled together the tags of all users with the same recency or semantic context values and calculated the tag reuse probability for both factors.

\begin{table}[t!]
	\small
  \setlength{\tabcolsep}{4.0pt}	
  \centering
    \begin{tabular}{l||cccc|c}
    \specialrule{.2em}{.1em}{.1em}
											Dataset			& $|U|$		& $|R|$			& $|T|$				& $|P|$							& $|P| / |R|$	\\\hline 
											Flickr						&	9,590		&	856,755		&	125,119		&	856,755										& 1.000					\\
											CiteULike						& 18,474  & 811,175		& 273,883		& 900,794 								& 1.110					\\
											BibSonomy	  				& 10,179  & 683,478		& 201,254			& 772,108 								& 1.129					\\
											Delicious					& 15,980 	& 963,741		& 184,012			& 1,447,267 							& 1.501					\\
											LastFM							& 1,892		& 12,522		& 9,748					& 71,062									& 5.674					\\
											MovieLens					& 4,009		& 7,601			& 15,238			& 55,484									& 7.299					\\
		\specialrule{.2em}{.1em}{.1em}								
    \end{tabular}
    \caption{Statistics of our datasets, where $|U|$ is the number of users, $|R|$ is the number of resources, $|T|$ is the number of distinct tags, $|P|$ is the number of posts and $|P| / |R|$ is the degree of narrowness.\vspace{-5mm}}	
  \label{tab:datasets}
\end{table}

\begin{figure*}[t!]
   \centering
   \subfloat[Flickr ($k$=.490, $R^2$=.532)]{ 
      \includegraphics[width=0.28\textwidth]{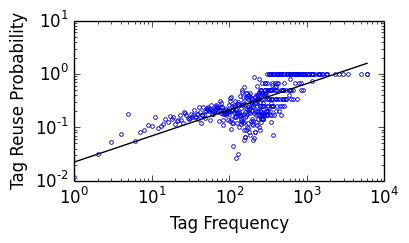}%
   }
	 \subfloat[Flickr ($k$=-1.074, $R^2$=.802)]{ 
      \includegraphics[width=0.28\textwidth]{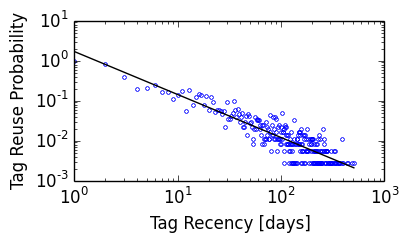}%
   }
   \subfloat[Flickr (-)]{ 
      \includegraphics[width=0.28\textwidth]{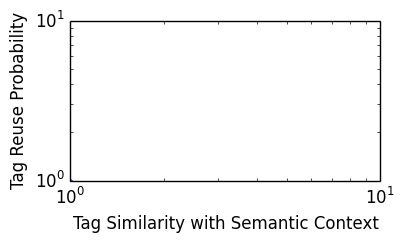}%
   }\\
	 \vspace{-4mm}
	 \subfloat[CiteULike ($k$=.708, $R^2$=.732)]{ 
      \includegraphics[width=0.28\textwidth]{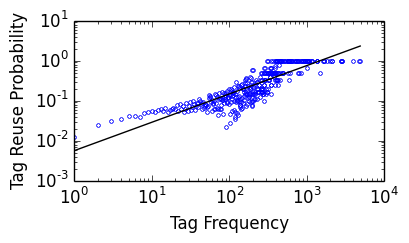}%
   }
	\subfloat[CiteULike ($k$=-.755, $R^2$=.672)]{ 
      \includegraphics[width=0.28\textwidth]{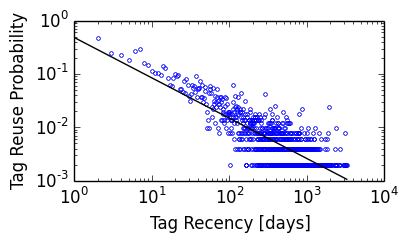}%
   }
		\subfloat[CiteULike ($k$=.565, $R^2$=.772)]{ 
      \includegraphics[width=0.28\textwidth]{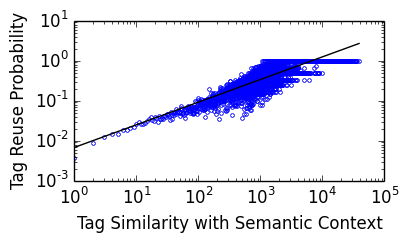}%
   }\\
		 \vspace{-4mm}
   \subfloat[BibSonomy ($k$=.733, $R^2$=.740)]{ 
      \includegraphics[width=0.28\textwidth]{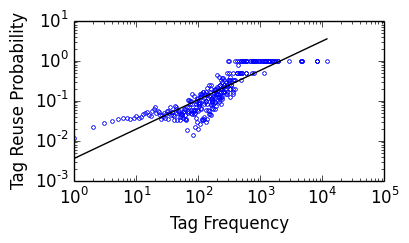}%
   }
	 \subfloat[BibSonomy ($k$=-.565, $R^2$=.495)]{ 
      \includegraphics[width=0.28\textwidth]{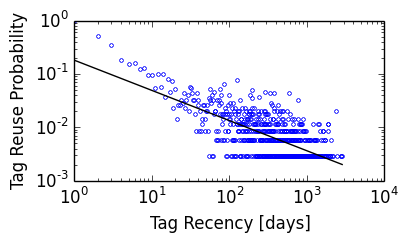}%
   }
	   \subfloat[BibSonomy ($k$=.492, $R^2$=.741)]{ 
      \includegraphics[width=0.28\textwidth]{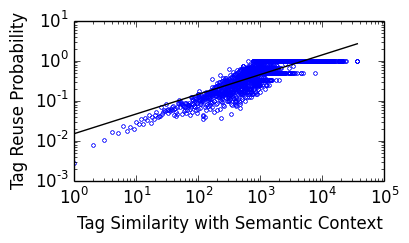}%
   }\\
		 \vspace{-4mm}
	 \subfloat[Delicious ($k$=.703, $R^2$=.700)]{ 
      \includegraphics[width=0.28\textwidth]{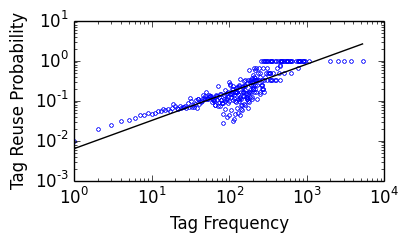}%
   }
		 \subfloat[Delicious ($k$=-1.415, $R^2$=.873)]{ 
      \includegraphics[width=0.28\textwidth]{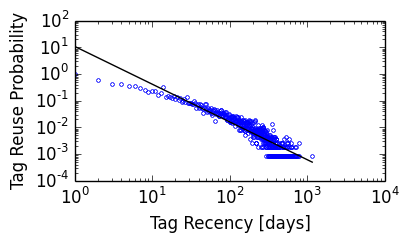}%
   }
		 \subfloat[Delicious ($k$=.728, $R^2$=.818)]{ 
      \includegraphics[width=0.28\textwidth]{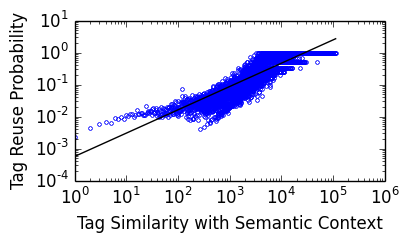}%
   }\\
		 \vspace{-4mm}
   \subfloat[LastFM ($k$=.427, $R^2$=.415)]{ 
      \includegraphics[width=0.28\textwidth]{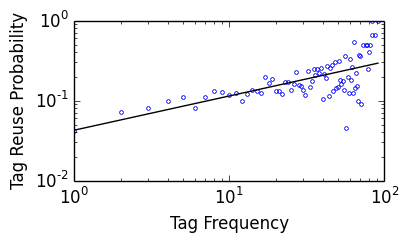}%
   }
	   \subfloat[LastFM ($k$=-.659, $R^2$=.342)]{ 
      \includegraphics[width=0.28\textwidth]{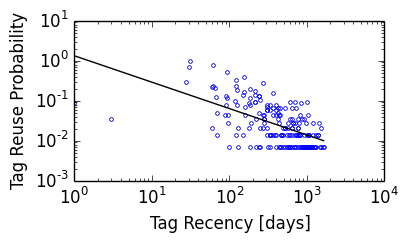}%
   }
	   \subfloat[LastFM ($k$=.558, $R^2$=.759)]{ 
      \includegraphics[width=0.28\textwidth]{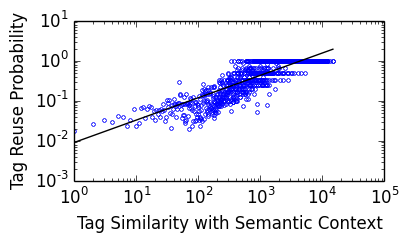}%
   }\\
		 \vspace{-4mm}
	 \subfloat[MovieLens ($k$=.838, $R^2$=.883)]{ 
      \includegraphics[width=0.28\textwidth]{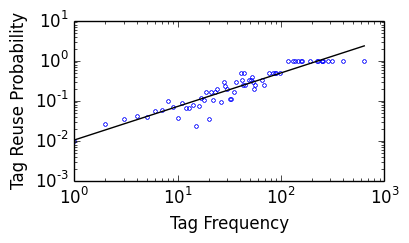}%
   }
		 \subfloat[MovieLens ($k$=-.243, $R^2$=.437)]{ 
      \includegraphics[width=0.28\textwidth]{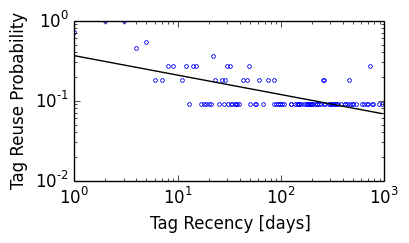}%
   }
		 \subfloat[MovieLens ($k$=.448, $R^2$=.685)]{ 
      \includegraphics[width=0.28\textwidth]{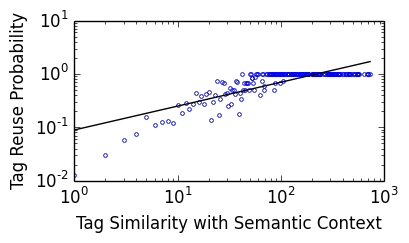}%
   }
   \caption{The influence of frequency, recency and semantic context on tag reuse in six social tagging systems (\textit{RQ1}). Clearly, all three factors positively influence a tag's probability of being reused: Slope $k$ > 0 for frequency, $k$ < 0 for recency and $k$ > 0 for semantic context. Furthermore, the generally high $R^2$ estimates of the linear regression indicate that large amounts of the data can be explained by a power function. \textit{Please note} that there is no semantic context in Flickr since users solely tag their own resources in this system.}
	 \label{fig:analysis}
\end{figure*}

\begin{table*}[t!]
  \small
  \setlength{\tabcolsep}{6.2pt}
  \centering
    \begin{tabular}{c|ll||ccc|ccc|cc}
    \specialrule{.2em}{.1em}{.1em}
									&				& 					& \multicolumn{3}{c|}{Individual factors}								& \multicolumn{3}{c|}{ Combination} & \multicolumn{2}{c}{Social}																												\\
			Folksonomy type 				& Dataset			& Metric 		& Frequency	& Recency	 	& SemCon	& GIRP 			  & BLL		  	& BLL$_{AC}$  													& FR			& PITF																	\\\hline 
			\multirow{2}{*}{\centering{Narrow}}	 & \multirow{2}{*}{\centering{Flickr}}																																																	                                    
									& $F_1$@5   & .371 			& .464							& -			&	.455				&	\textbf{.470}			&	.470						& .365		& .350												\\						
						&			&	nDCG@10		& .569 			& .702							& -			&	.686 				&	\textbf{.711}     &	.711		        & .561    & .535                  				\\\hline                                  
			\multirow{6}{*}{\centering{Mixed}}	 & \multirow{2}{*}{\centering{CiteULike}}																																					                                                            
									& $F_1$@5   & .231 			& .236 							& .041 	& .243				&	.254							&	\textbf{.259}						&	.250		&	.178					\\
						&			&	nDCG@10		& .367 			& .385 							& .069 	& .394				&	.413 							&	\textbf{.422}      			&	.392		&	.294			\\\cline{2-11}    
			 & \multirow{2}{*}{\centering{BibSonomy}}																																												                                              
									& $F_1$@5   & .253 			& .252 							& .063  & .262				&	.269							&	\textbf{.280}						&	.279		&	.215			\\
						&			&	nDCG@10		& .371 			& .368 							& .090  & .386				&	.396 							&	\textbf{.409}      			&	.408		&	.327				\\\cline{2-11}      
			 & \multirow{2}{*}{\centering{Delicious}}																																								                                                      
									& $F_1$@5   & .173 					& .179 					& .108 				& .190							&	.203	&	\textbf{.243}						&	.196		&	.199			\\
						&			&	nDCG@10		& .267 					& .287 					& .158 				& .298							&	.318 	&	\textbf{.374}      			&	.292		&	.302					\\\hline     
			 \multirow{4}{*}{\centering{Broad}}	& \multirow{2}{*}{\centering{LastFM}}																																														                                          
									& $F_1$@5   & .193 					& .189 					&  .202				& .198							&	.202	&	.251										&	.270		&	\textbf{.276}						\\
						&			&	nDCG@10		& .292 					& .293 					&  .302				& .303							&	.313 	&	.375    								& .399 		& \textbf{.414}							\\\cline{2-11}      		
			& \multirow{2}{*}{\centering{MovieLens}}																																																										                  
									& $F_1$@5   & .077 					& .076 					& .077 				& .077							&	.079	&	.086										&	.153		&	\textbf{.156}						\\
						&			&	nDCG@10		& .177 					& .183 					& .176 				& .177							&	.187 	&	.203    								& .319 		& \textbf{.324}				\\
							
		\specialrule{.2em}{.1em}{.1em}
    \end{tabular}
		\caption{Prediction accuracy results of algorithms that (i) reflect the individual factors of frequency, recency and semantic context, (ii) combine these factors, and (iii) utilize social influences (\textit{RQ2}). We see not only that all three individual factors can be exploited to efficiently predict a user's tags but also that the efficacy of the algorithms depends on the folksonomy type (i.e., narrow, mixed, broad) of the given system.\vspace{-4mm}}
		\label{tab:results}		
\end{table*}

\subsection{Results} \label{sec:emp_results}
In Figure \ref{fig:analysis}, we plotted the tag reuse probability over (i) tag frequency, (ii) tag recency (in days), and (iii) tag similarity with the current semantic context on a log-log scale for our six datasets. We also provided the $k$ (i.e., the slope) and $R^2$ (i.e., the determination coefficient) estimates of the linear regression on the data. Specifically, we use the sign of $k$ to determine how tag reuse is influenced by the factors and $R^2$ to check if the data follows a power function. Across all six datasets, we can make three main observations:\vspace{-1mm} %
\begin{enumerate}
\item The \textit{more frequently} a tag was used in the past ($k$ > 0), the higher its reuse probability is.\vspace{-1mm}
\item The \textit{more recently} a tag was used in the past ($k$ < 0), the higher its reuse probability is.\vspace{-1mm}
\item The \textit{more similar} a tag is to tags in the \textit{current semantic context} ($k$ > 0), the higher its reuse probability is.
\end{enumerate}

As such, all three factors (i.e., frequency, recency and semantic context) have a positive influence on a tag's probability of being reused. Furthermore, the generally high $R^2$ estimates in the log-log scaled plots indicate that large amounts of the data can be explained by a power function, as also indicated by the first component of the activation equation \cite{anderson1991reflections}. Thus, the first research question of our work (\textit{RQ1}) can be answered affirmatively. In order to better understand the (individual) influence of each of the factors for predicting tags given a specific folksonomy type (\textit{RQ2}), we conducted a prediction study described in the next section.%

\section{Prediction Study} \label{sec:pred}
In this section, we present the algorithms, methodology and results of our prediction study, which was conducted to address our second research question (\textit{RQ2}). Specifically, we want to establish to what extent the factors of frequency, recency and semantic context can be exploited individually to efficiently predict a user's tag reuse given a specific folksonomy type of a social tagging system.

\subsection{Algorithms} \label{sec:algorithms}
In terms of the compared algorithms, we utilize approaches that reflect the three factors individually, approaches that combine these factors and approaches that incorporate social influences (e.g., via related tags of other users).

\vspace{2mm} \noindent \textbf{Individual factors.} To account for tag frequency, we utilize the Most Popular Tags (MP$_u$) approach, which ranks the tags of a user based on their usage frequency \cite{jaschke2007tag}. To predict tags solely based on tag recency, we rank the tags of a user by the timestamp of their last usage. Since the third factor of interest, the semantic context \textit{SemCon}, is represented by the second component of the activation equation, we recommend the tags that highly co-occurred with tags already assigned to the currently tagged resource \cite{springer_bllac}.

\vspace{2mm} \noindent \textbf{Combination of factors.} We utilize three algorithms based on combinations of the factors, which enables us to analyze if the performance of the individual factors can be improved when they are combined in the form of hybrid approaches. The first one, \textit{GIRP} \cite{zhang2012integrating}, integrates frequency and recency using an approach that models the effect of time (i.e., recency) via an exponential function. The second one, \textit{BLL} \cite{www_bll}, implements the first component of the activation equation and models the effect of time via a power function as suggested by \cite{anderson1991reflections}. The third algorithm in this respect, \textit{BLL$_{AC}$} \cite{springer_bllac}, is the full implementation of the model, accounting for all three factors (frequency, recency and semantic context).

\vspace{2mm} \noindent \textbf{Social influences.} To compare these methods that aim to predict a user's individual tag reuse with approaches that also integrate social influences (e.g., related tags of other users), we incorporate two well-known algorithms from tag recommender research in our study. The first one, FolkRank (\textit{FR}) \cite{hotho2006folkrank}, is an extension of Google's PageRank approach to iteratively rank the entities in folksonomies. The second one, Pairwise Interaction Tensor Factorization (\textit{PITF}) \cite{rendle2010pairwise}, is based on factorization machines and has become one of the most successful methods for recommending tags. For both algorithms, we use the same parameter settings as in \cite{recsys_eval} (i.e., we set the preference vector weighting $d$ for FR to .7 and the dimension of factorization $\lambda$ for PITF to 256).

\subsection{Methodology} \label{sec:metrics}
We conducted our prediction study on the same datasets and training/test set splits as in our empirical study (see Section \ref{sec:emp_meth}). This allowed us to train the algorithms based on the posts in the training set and compare the top-$10$ tags that an algorithm predicted for user $u$ and resource $r$ of a post in the test set with the set of relevant tags in this test set post \cite{jaschke2007tag,jaschke2008tag}. Based on that, we computed two prediction accuracy metrics known from research on information retrieval and recommender systems. Specifically, we report the F1-score ($F_1$@5) for the top-$5$ predicted tags\footnote{$F_1$@5 was also used as the main metric in PKDD'09: \url{http://www.kde.cs.uni-kassel.de/ws/dc09/evaluation}} and the ranking-aware metric Normalized Discounted Cumulative Gain (nDCG@10) for the top-$10$ predicted tags \cite{lipczak2012hybrid}.

To ensure the reproducibility of our results, we conducted this study via the open-source tag recommender evaluation and benchmarking framework \textit{TagRec} \cite{ht_tagrec}, which can be freely downloaded from GitHub\footnote{\url{https://github.com/learning-layers/TagRec/}} for scientific purposes.

\subsection{Results} \label{sec:pred_results}
In Table \ref{tab:results}, we present the results of our prediction study indicated by the $F_1$@5 and nDCG@10 metrics. Across all six datasets and both metrics, we observe three patterns of results based on the folksonomy type of the tagging systems.

\vspace{2mm} \noindent \textbf{Narrow.} In the narrow folksonomy Flickr, the semantic context has no influence since users are solely tagging their own images. The results show that frequency and especially recency can be exploited to efficiently predict a user's tag reuse in this narrow setting. Furthermore, these factors even outperform FR and PITF, the two algorithms that also utilize social influences by means of other users' tags. When combining frequency and recency, we see that the accuracy of the strong recency factor can only be slightly improved in the case of BLL, which models the time component via a power function, and even decreases in the case of GIRP, which builds on an exponential temporal decay function.

\vspace{2mm} \noindent \textbf{Mixed.} For the mixed folksonomies CiteULike, BibSonomy and Delicious, we observe a good performance for the factors of frequency and recency, and an average one for the semantic context. Additionally, the results suggest that a combination of all three factors in the form of BLL$_{AC}$ provides the highest accuracy estimates and outperforms FR and PITF. Again, BLL (and thus, the power function) is apparently better suited to combine frequency and recency than GIRP (and thus, the exponential function).

\vspace{2mm} \noindent \textbf{Broad.} Interestingly, the algorithms in the broad folksonomies LastFM and MovieLens had a completely different behavior. In these datasets, since there are a lot of tags assigned by other users to the currently tagged resource, the semantic context is a much more important factor for predicting tags than in the narrow and mixed settings. Similarly to the narrow case, the combination of the factors only slightly improves the accuracy of the individual factors. Due to a high number of average posts per resource (5.674 for LastFM and 7.299 for MovieLens -- see Table \ref{tab:datasets}), FR and PITF, which utilize related tags of other users as well, provide the best results in the broad setting.

\vspace{2mm} \noindent \textbf{Summary.} Our results are summarized in Table \ref{tab:discussion}. Overall, they confirm that the factors of frequency, recency and semantic context can be exploited to efficiently predict a user's tag reuse. Which factor to consider, however, strongly depends on the folksonomy type of the given social tagging system. As such, the second research question of our work (\textit{RQ2}) can also be answered affirmatively.

\section{Conclusion and Future Work} \label{sec:conc}
In this paper, we analyzed the influence of frequency, recency and semantic context on the reuse of tags in social tagging systems. To that end, we conducted an empirical study and a prediction study on datasets gathered from the six social tagging systems Flickr, CiteULike, BibSonomy, LastFM and MovieLens. The empirical study aimed to answer our first research question of this work (\textit{RQ1}) and determine how the factors of frequency, recency and semantic context influence a tag's probability of being reused. The three main findings of this analysis are: (i) the \textit{more frequently} a tag was used in the past, the higher its reuse probability is, (ii) the \textit{more recently} a tag was used in the past, the higher its reuse probability is, and (iii) the \textit{more similar} a tag is to tags in the \textit{current semantic context}, the higher its reuse probability is. This confirms that all three factors (i.e., frequency, recency and semantic context) have a positive influence on a tag's probability of being reused.

Our prediction study was designed to determine to what extent these three factors can be exploited to efficiently predict a user's tag reuse given a specific folksonomy type (\textit{RQ2}). With that regard, we applied not only algorithms that reflect these three factors individually (\textit{Frequency}, \textit{Recency} and \textit{SemCon}) but also algorithms that combine these factors (\textit{GIRP}, \textit{BLL} and \textit{BLL$_{AC}$}) and incorporate social influences, i.e., related tags of other users (\textit{FR} and \textit{PITF}). We observed three patterns of results based on the folksonomy type of the datasets, which are summarized in Table \ref{tab:discussion}. In the narrow case, the factor of recency is the most important one, whereas in the mixed case all three factors highly contribute to the prediction accuracy and thus, the combination of these factors (\textit{Comb}) provides the best results. Finally, in case of the broad folksonomies, the semantic context becomes a much more important factor than in the other two settings. Furthermore, the best results in this setting are obtained for PITF, which utilizes social influences (\textit{Social}), such as related tags of other users, as well.

Overall, our results show that frequency, recency and semantic context positively influence the reuse probability of tags in all six datasets (\textit{RQ1}) and that the efficacy of these factors in terms of predicting a user's tags depends on the folksonomy type of the given social tagging system (\textit{RQ2}). Additionally, with our findings summarized in Table \ref{tab:discussion}, we provide a transparent overview of the factors that influence the prediction of tags based on the given folksonomy type. We believe that our findings may be a guideline for researchers and developers in the area of tag-based recommender systems, helping them to choose the correct prediction method for a given social tagging environment.\vspace{2mm}

\begin{table}[t!]
	\small
  \setlength{\tabcolsep}{2.2pt}	
  \centering
    \begin{tabular}{c||ccc|c|c}
    \specialrule{.2em}{.1em}{.1em}										
											Folksonomy type		& Frequency		& Recency			& SemCon		& Comb		& Social	\\\hline 
											Narrow						&	+/-					&	+						&	-							&	+/-			& -					\\
											Mixed							& +					 	& +						& +/-						&	+				& +/-					\\
											Broad							& +/-					& +/-					& +							&	+/-			& +					\\
		\specialrule{.2em}{.1em}{.1em}								
    \end{tabular}
    \caption{Summary of our prediction accuracy results showing the performance of the algorithms based on the given folksonomy type. \textit{Please note} that ``+'' indicates a good performance, ``+/-'' indicates an average performance and ``-'' indicates a poor performance of a factor/an approach in a specific setting.\vspace{-5mm}}	
  \label{tab:discussion}
\end{table}

\noindent \textbf{Limitations and future work.} One limitation of this study is that it is mainly focused on the factors that influence the \textit{individual} reuse of tags. Thus, for future work, we would like to extend this by also analyzing the influence of the reuse (i.e., imitation) of other users' tags and study how this \textit{social influence} effects tag predictions. Furthermore, we plan to expand our model of the \textit{semantic context}, since to date, we have analyzed it solely based on the tags that were already assigned to the currently tagged resource. In this respect, we want to incorporate other resource-dependent information, such as the resource title or content, as well.

Finally, we plan to enhance our guideline with results for other types of social systems that utilize the concept of tags. To that end, we would like to determine the influence of frequency, recency and semantic context in systems that incorporate \textit{hashtags} such as Twitter, Facebook and Instagram. At the same time, this would allow us to verify that our findings can be generalized to various types of tags.\vspace{2mm}

\noindent \textbf{Acknowledgments.} The authors would like to thank Paul Seitlinger, Tobias Ley, Ilire Hasani-Mavriqi and Emanuel Lacic for valuable comments on this work. This work is funded by the Know-Center and the EU-IP Learning Layers.%

\bibliographystyle{abbrv}

\end{document}